\documentstyle[12pt]{article}

\begin{document}

\baselineskip 24pt

\title{\large{\bf
 Exact Results And Soft Breaking Masses In Supersymmetric Gauge
Theory }}

\author{
Nick Evans\thanks{nick@zen.physics.yale.edu},
Stephen D.H.~Hsu\thanks{hsu@hsunext.physics.yale.edu}, \\
Myckola Schwetz\thanks{ms@genesis2.physics.yale.edu}, \/ and
Stephen B.~Selipsky\thanks{stephen@genesis1.physics.yale.edu}
\\ \\ Department of Physics \\ Yale University
\\ New Haven, CT 06520-8120 }

\date{August 1, 1995}

\maketitle

\begin{picture}(0,0)(0,0)
\put(350,300){YCTP-P11-95}
\put(350,315){hep-th/9508002}
\end{picture}
\vspace{-24pt}

\begin{abstract}
We give an explicit formalism connecting softly broken supersymmetric
gauge theories (with QCD as one limit) to $N=2$ and $N=1$
supersymmetric theories possessing exact solutions, using spurion
fields to embed these models in an enlarged $N=1$ model.  The
functional forms of effective Lagrangian terms resulting from soft
supersymmetry breaking are constrained by the symmetries of the
enlarged model, although not well enough to fully determine the
vacuum structure of generic softly broken models.  Nevertheless by
perturbing the exact $N=1$ model results with sufficiently small
soft breaking masses, we show that there exist nonsupersymmetric models
that exhibit monopole condensation and confinement in the same modes
as the $N=1$ case.
\end{abstract}

\section{Introduction} \label{sec-intro}

Remarkable, exact results on the vacuum structure of four dimensional
$N=2$ supersymmetric quantum field theories with an $SU(2)$ gauge
symmetry and matter fields in the fundamental representation have
recently been obtained \cite{SW}, with generalizations to $SU(N_c)$
gauge groups \cite{Ncolor}.  In these models, an $N=1$ supersymmetry
preserving mass perturbation for the adjoint superfield leads to
monopole condensation, confinement and chiral symmetry breaking.
There has been much interest in how such exact supersymmetric
results generalize to nonsupersymmetric gauge theories
\cite{othersusybreaking}.

In this paper we show how $N=2$, $N=1$ and softly broken
supersymmetric models can all be embedded in a single enlarged $N=1$
model in which coupling constants are treated as ``spurion'' fields.
The effective superpotential of the enlarged model can be exactly
determined from its symmetries and the $N=2$ limit.  In the $N=2$
limit the $D$-terms are also exactly determined, but when
supersymmetry is broken the potential and superpotential each receive
contributions from additional $D$-terms of the enlarged model, which
vanish in the $N=2$ limit.  Although constrained by the symmetries of
the model, these cannot be fully determined.  If the $N=2$ symmetry
is broken solely by soft breaking masses, these unknown functions
typically are necessary to determine details of the vacuum structure.
The exact superpotential does however include all the lowest-dimension
gauge interaction terms, and continues to have singularity
structure\footnote{
 Strictly speaking there is no longer a moduli space after
 breaking $N=2$ supersymmetry.  However, we can still discuss
 monodromies in the configuration space of the theory, by means
 of an external source used to traverse a closed loop.}
consistent with points at which monopoles become massless.
We can thus confirm that when the soft breaking masses are
perturbations to an $N=1$ preserving mass, the vacuum remains
essentially that of the $N=1$ model.

An important step in obtaining the exact results of \cite{SW} is to
treat the couplings in the $N=2$ models as spurion chiral
superfields in an enlarged $N=1$ model, which fail to propagate if
their kinetic term coefficients are taken to infinity \cite{natiral}.
Their scalar component vacuum expectation values (vevs) can thus be
frozen at any chosen values.  The $F$-terms of the Wilsonian effective
Lagrangian of a supersymmetric model \cite{natiral,SV,CERN} must be
holomorphic in the fields as well as invariant under the gauge and
global symmetries of the model (assuming a supersymmetric
regularization scheme).  The superpotential of the enlarged model is
therefore holomorphic in the couplings.  A recent paper \cite{yale1}
showed that soft supersymmetry breaking interactions and mass terms
can be introduced in supersymmetric models without altering the
holomorphic constraint on the $F$-terms.  The spurion fields can be
coupled to a sector that generates supersymmetry breaking expectation
values for the spurion field $f$-components\footnote{
 Our superfield notation \cite{WB} is $\Phi = a_\Phi
 + \sqrt{2}\, \theta \psi_\Phi + \theta \theta f_\Phi$,
 with $\Phi^\dagger\Phi\big|_D \equiv
 \int d^2\theta\ d^2{\bar\theta}\ \Phi^\dagger\Phi = |f_\Phi|^2,\
 \Phi\big|_F \equiv \int d^2\theta\ \Phi = f_\Phi,\
 \Phi\big|_A \equiv a_\Phi$, etc.}.
Freezing out the spurions generates soft supersymmetry breaking masses
and interactions in the embedded model.

For large vevs of the adjoint scalar field $a_\Phi$ in the models we
study, the $SU(N_c)$ gauge symmetry is
broken to a subgroup of weakly interacting $U(1)$ gauge symmetries.
It has been argued \cite{SW} that the $N=2$ theory remains in the
Coulomb phase even for small scalar vev where the model is
strongly coupled.  One might have expected the pure $SU(N_c)$
dynamics to operate unimpeded when the scalar vev was much smaller than
the strong interaction scale $\Lambda$, and the low energy theory
to exist in a confining phase with all the gauge bosons confined
within glueballs.  However, the solutions' self-consistency
\cite{SW} suggests that for the special choice of couplings that
gives the $N=2$ model, the quantum effects that cause condensation
and confinement exhibit special cancellations, leading to the
surprising low energy dynamics of a strongly coupled Coulomb phase
with massless fermions, monopoles and dyons.

The solutions' consistency with our expectations for how these
theories ought to behave helps confirm that higher dimension terms
do not qualitatively change the solutions.  Following the analysis
\cite{SW}, which does not determine operators with more than two
derivatives (${\cal O}(p^2)$) or four fermions, we will assume that
the gauge particle dynamics are essentially controlled by the
effective operators of lowest dimension, at sufficiently low energy
scales.  The model's symmetries permit higher dimension operators
containing for example powers of $|WW/\Lambda^3|^2$, and (by $N=2$
supersymmetry) also containing superderivatives and powers of
$\Phi/\Lambda$.  Since the scalar component of $\Phi$ is pinned at
$\langle a_\Phi \rangle \sim \Lambda$ when the supersymmetry is broken
to $N=1$, these terms are not at first sight negligible.  However,
since they must be related by $N=2$ supersymmetry to the corresponding
higher derivative $W$ terms \cite{mans}, they should also be suppressed
in the infrared regime.  Thus despite these caveats, a tractable
electric--magnetic duality is useful in determining the theory's
vacuum structure.

We expect that introducing $N=2$ breaking masses destroys
cancellations in the dynamics below the scale of the breaking mass,
and confines the $U(1)$ gauge interaction below that scale (via the
Higgs mechanism in the dual theory).  Provided that breaking masses
are sufficiently small relative to the scale $\Lambda$ of the strong
interactions, the theory will continue to be described by a $U(1)$
gauge theory between the $SU(2)$ breaking scale parameterized by
$\langle a_\Phi \rangle$, pinned at $\sim \Lambda$, and the breaking
mass scale.  Softly broken $N=1$ and $N=0$ models, when close in
parameter space to the $N=2$ model, are therefore anticipated to
occupy the same phase.

The paper is organized as follows.  Section~\ref{sec-su2} explicitly
describes the class of models we study.  We allow the models' coupling
constants to be chiral superfield spurions, whose values are
eventually frozen in all of spacetime \cite{natiral,yale1}.  We then
review the analysis that leads to the Seiberg--Witten ansatz for the
vacuum structure for $SU(2)$ gauge symmetry, and monopole condensation
in $N=1$ models close in parameter space to the $N=2$ model.  Allowing
soft supersymmetry breaking terms in our Lagrangian via nonvanishing
spurion field $f$-components, we show that to lowest order in the soft
breakings, gauge kinetic terms are unchanged by the introduction of
soft masses.  However, the potential minimum in these models typically
depends upon insufficiently constrained contributions from $D$-terms.
Section~\ref{sec-matter} discusses obtaining similar results in $SU(N_c)$
gauge theories or with matter fields in the fundamental representation,
and reaching QCD by continuous interpolation from models of this type.

\newpage
\section{Supersymmetric $SU(2)$} \label{sec-su2}
\subsection{An enlarged model} \label{sec-2.1}

Consider an $N=1$ supersymmetric $SU(2)$ gauge theory, containing one
matter chiral superfield in the adjoint representation.  Promoting
the coupling constants into chiral superfields (including a $D$-term
normalization, $K$, which we shall use to generate a squark mass)
\cite{yale1} yields the Lagrangian
\begin{equation} \label{smallLagrangian}
 \begin{array}{c}
 {\cal L} =
 {1\over 4\pi}{\rm Im}\left( {1\over 2}\tau_0 W_\alpha W^\alpha
                             \Big|_F
 \ +\ (\tau_0 + K^\dagger K) \Phi^\dagger e^V \Phi\Big|_D \right)
 \ +\ m \Phi^2\Big|_F \ +\ {\rm h.c.}\\ \\
 \ +\ \Lambda_m^2 \left(m^\dagger m\Big|_D + \beta_m m\Big|_F
    + {\rm h.c.}\right)
 \ +\ \Lambda_\tau^2 \left( \tau_0^\dagger \tau_0 \Big|_D
    + \beta_\tau \tau_0 \Big|_F  + {\rm h.c.}\right)\\ \\
 \ +\ \Lambda_K^2 \left( K^\dagger K\Big|_D + \beta_K K\Big|_F
 + {\rm h.c.}\right)
 \end{array}
\end{equation}
(note that $\Lambda_m$ and $K$ are dimensionless).
The spurion fields $i = \{m, \tau_0, K\}$ do not propagate in the
limit $\Lambda_i \rightarrow \infty$; we can fix their scalar
components $a_i$ at any expectation values we choose, and replace
$f_i = -\beta_i^\dagger + {\cal O}(1/\Lambda_i^2)$.
[An equivalent approach normalizes the spurion $D$-terms
conventionally, placing the $\Lambda_i$ suppression on
the spurion-to-physical-field coupling operators; for example
writing $m^\dagger m\Big|_D + (m/\Lambda_m)\Phi^2\Big|_F$.  The
spurion vevs are then taken proportional to $\Lambda_i$, so their
classical and quantum evolutions are still relatively suppressed.]

In any case, if we choose $\langle a_{\tau_0} \rangle\neq 0$
and $\left(\langle a_m\rangle\ ,\ \langle a_{K} \rangle\ ,\
\langle f_i \rangle\ \right) \ =\ 0$,
the model reduces to $N=2$ supersymmetric $SU(2)$, studied in
\cite{SW}, whose bare Lagrangian is
\begin{equation} \label{neqone}
 {\cal L}_{N=2}\ =\ {1\over 4\pi}{\rm Im} \langle a_{\tau_0}\rangle
 \left( {1\over 2} W_\alpha W^\alpha \Big|_F
 \ +\ \Phi^{\dagger} e^V \Phi\Big|_D \right)\ .
\end{equation}
In this notation, we will review the derivation \cite{SW} of the
$N=2$ model effective potential and its generalization to the
$N=1$ case of $\langle a_m \rangle \neq 0$.  Nonzero values of
$\langle f_{\tau_0} \rangle$ and $\langle f_K\rangle$ generate
supersymmetry breaking mass terms for gauginos and squarks
respectively.
In the limit $\langle f_{\tau_0,K} \rangle \rightarrow\infty$
the gauginos and squarks decouple from the low energy theory,
leaving only $SU(2)$ gauge bosons and adjoint fermions.

The model has an anomaly-free $U(1)_R$ global symmetry, whose charge
assignments
\begin{equation} \label{U1Rcharges}
 \matrix{
 \theta   && +1  \cr
 W_\alpha && +1  \cr
 \tau_0   &&  0  \cr
 \Phi     &&  0  \cr
 m        && +2  \cr
 K        && {\rm arbitrary} }
\end{equation}
significantly constrain the terms which can appear in the low energy
effective Lagrangian.  Allowable $D$-term operators have net $R$-charge
of zero, whereas $F$-term operators must have net $R$-charge of $+2$.
The anomaly breaks a second $SU(2)_R$ symmetry of the embedded $N=2$
model to a discrete symmetry; in the $SU(2)$ gauge case this gives
the ${\bf Z_2}$ symmetry.

A further constraint applies to the spurion sources $\beta_i$, which
may be treated as decoupled chiral superfields in their own right.
Coupled only linearly to the $f$-components of $m,\tau$ and
$K$, they do not contribute perturbatively to 1PI diagrams.
Moreover, since scalar components $a_{\beta_i}$ do not couple at all
to other fields in the bare potential, they will only appear
suppressed by powers of $\Lambda_i$ in induced terms in the
effective Lagrangian.
The only dependence in the effective theory on $\beta_i$ vevs
thus occurs indirectly, through the $f$-component vevs $f_i$ of
coupling-field spurions, induced by $a_{\beta_i}$.

\subsection{The low energy effective theory in the $N=2$ limit}
\label{sec-2.2}

To fix our notation we briefly review the analysis \cite{SW} relevant
to the $N=2$ model in (\ref{neqone}).  The form of the ${\cal O}(p^2)$
effective Lagrangian's superpotential can be deduced from its holomorphic
properties as described below.  In the $N=2$ limit the effective
Lagrangian's $D$-terms are then determined by $N=2$ supersymmetry,
giving the effective Lagrangian as
\begin{equation} \label{N=2lowE}
 {1\over 4\pi} {\rm Im}\left\{ {\partial{\cal F}(A) \over \partial A}
 \bar A\Bigg|_D
 \ +\ {1\over 2} {\partial^2{\cal F}(A) \over \partial A^2}
  W_{\alpha} W^{\alpha}\Bigg|_F \right\} \ ,
\end{equation}
where $A$ is the $N=1$ chiral multiplet and ${\cal F}$ the
``prepotential'' \cite{SW,prepotential}.  The exact solution for the
effective Lagrangian of the $N=2$ model determines the effective
Lagrangian of the enlarged model, and hence of the models with soft
supersymmetry breaking masses, up to terms compatible with the
symmetries of the enlarged model that vanish in the $N=2$ limit.
We shall discuss these extra terms in the following sections.

In the $N=2$ model the classical potential for $a_\Phi$,
$V_{cl} = {\rm Tr}([a_\Phi, a_\Phi^\dagger]^2)$,
is minimized (and vanishes) when $a_\Phi$ has any diagonal complex vev.
This breaks $SU(2)$ down to $U(1)$ and yields a classical moduli space,
parameterized by the gauge invariant superfield $U \equiv {\rm Tr}[\Phi^2]$.
We assume that at low energies the quantum theory also remains in the
Coulomb phase, so that the particle content is restricted to the
$U(1)$ gauge superfield $W_\alpha$ and the massless matter fields.

In the Coulomb phase at arbitrarily low scales, where we need consider
only the lowest dimension terms in the Lagrangian, the effective action
is simply quadratic in the gauge field, and a duality transformation is
conveniently derived \cite{SW}
by considering a functional integral over the gauge superfield $W$.
Imposing the condition Im$({\cal D} W) = 0$ via a Lagrange multiplier
$V$, incorporated by a new dual gauge field $W_D = i{\cal D}V$, we can
complete the square and perform the Gaussian integral, to obtain an
equivalent theory for the dual field with $\tau_D = -1/\tau$:
\begin{equation} 
 {\cal Z} \sim \int{\cal D}[W_D]\ \exp{\left({i\over 8\pi}
  {\rm Im} \int \tau_D W_D^\alpha W_{D \alpha}\Big|_F \right)}\ .
\end{equation}
In order to maintain the explicit $N=2$ supersymmetry of (\ref{N=2lowE})
in the dual theory we must rewrite the matter $D$-terms
\begin{equation} \label{dualprepotential}
 {\partial {\cal F}(A) \over \partial A} \bar{A} \ =\
 {\partial {\cal F_D}(A_D) \over \partial A_D} \bar{A}_D \ ,
\end{equation}
with $\tau \ =\ \partial^2 {\cal F}(A)/ \partial A^2$ and
$\tau_D \ =\ \partial^2 {\cal F}_D(A_D)/ \partial A_D^2$.  Thus
\begin{equation} \label{taudfda}
 A_D \ =\ {\partial {\cal F(A)} \over \partial A}\ , \qquad
 \tau \ =\ {\partial A_D \over \partial A}\ .
\end{equation}
This manipulation is independent of whether $\tau$ is a coupling
constant or, as in our enlarged model, a chiral superfield itself.
Duality therefore continues to hold at low energies even after
supersymmetry is broken by for example $\langle f_\tau \rangle \neq 0$,
induced by $f_{\tau_0} \neq 0$; in the dual theory, the trivially related
$\langle f_{\tau_D} \rangle$ drives supersymmetry breaking.

The theory is also invariant under shifts by integer $n$ in the real part
of $\tau$, since these correspond to unobservable shifts in the $\theta$
angle.  Writing $\tau$ as the ratio of the two components of a vector
allows us to recognize the duality and $\theta$ angle transformations
\begin{equation} \label{tautransform}
 \left( \begin{array}{cc} a & b\\ c & d \end{array} \right)
 \left( \begin{array}{c} \tau \\ 1 \end{array} \right)
\end{equation}
as generating the group $SL(2,Z)$.

Now consider the effective theory for scales much less than $U$.
Its $F$-terms must be holomorphic and invariant under the global
$U(1)_R$ symmetry of (\ref{U1Rcharges}), and therefore of the form
\begin{equation}
 {\cal L}_{\rm gauge} \ =\ {1\over 8\pi}{\rm Im} \left[
     \tau(\tau_0,U)\, W_\alpha W^\alpha \Big|_F \right] \ .
\end{equation}
If $\tau$ were everywhere analytic in $U$, Im$(\tau)$ would be a harmonic
function, which would be unbounded below, resulting in an imaginary gauge
coupling.  To avoid this $\tau$ must have singularities at finite values
of $U$, $U_i$, presumably due to composite states driven massless by
strong interactions.  Thus the dual theory is weakly coupled near the
singularities.  The composite (monopole or dyon) fields are light near
the singular points and must be included in the effective theory.
We therefore include ``hypermultiplet'' terms compatible with the
symmetries in the effective Lagrangian:
\begin{equation} \label{Lmonopole}
 {\cal L}_{\rm dyon} \ =\ M^\dagger e^V M \Big|_D
 \ +\ {\tilde M}^\dagger e^{-V} {\tilde M} \Big|_D
 \ +\ \sqrt{2} A_D M {\tilde M} \Big|_F \  \ +\ {\rm h.c.}\ ,
\end{equation}
where $M$ and ${\tilde M}$ are the two $N=1$ chiral multiplets of an $N=2$
hypermultiplet, and $A_D \sim (U-U_i)$ determines the bound states' mass.

The simplest possibility for $\tau$, consistent with its weak-coupling limit
at large $U$ and with ${\bf Z_2}$ symmetry $U \leftrightarrow -U$, is a pair
of logarithmic singular points at each of which a single composite state
becomes massless.  An $SL(2,Z)$ invariant function with two such singularities
is determined up to scalings by its behavior there and at infinity, and can be
interpreted as the modular parameter of a torus
(with $\tau$ the ratio of its two periods).
The torus described by $\tau$ corresponds to a cubic elliptic curve with roots
$e_1, e_2, e_3$,
\begin{equation} \label{ellcurve}
 y^2 \ =\ 4 (x - e_1) (x - e_2) (x - e_3) \ .
\end{equation}
Up to $SL(2,Z)$ transformations, $\tau$ can be written
\begin{equation} \label{tau}
 \tau \ =\ {\partial A_D \over \partial A} \ =\
 {\int_{e_1}^{e_2} {dx / y} \over \int_{e_2}^{e_3} {dx / y}}\ .
\end{equation}

Seiberg and Witten showed that if the dyons are respectively a $(0,1)$
magnetic monopole and a $(1,-1)$ dyon, then one obtains a function $\tau$
consistent
with the one loop beta function at $U \rightarrow \infty$:
 $\tau(U\rightarrow\infty) \sim -(i/\pi) \ln U$.
The charges are redefinable by circuiting infinity, which induces
 $(n_m,n_e)\rightarrow (-n_m,-n_e - 2 n_m)$.
Furthermore, changing the $\theta$ angle of $\tau$ by $2\pi$ simply
switches the labelling of the monopole and dyon, leaving the two singular
points $U_i$ physically equivalent, as the ${\bf Z_2}$ symmetry requires.
For the pure gauge $N=2$ $SU(2)$ model, these boundary conditions
correspond to an elliptic curve
\begin{equation} \label{ellipticSU2}
 y^2 \ =\ (x - U_i)(x + U_i)(x - U) \ ,
\end{equation}
where $U_2 = -U_1$ by the ${\bf Z_2}$ symmetry.  The period integrals are
\begin{eqnarray} \label{aadintegrals}
 A &\ =\ & {\sqrt{2}\over\pi} \int^{U_i}_{-U_i}
 {dx \sqrt{x - U} \over \sqrt{x^2-U_i^2}} \\
 && \nonumber \\ A_D &\ =\ & {\sqrt{2}\over\pi} \int_{U_i}^U
 {dx \sqrt{x-U} \over \sqrt{x^2-U_i^2}}\ .
\end{eqnarray}
These integrals can be expanded in powers of $z \equiv (U - U_i)/U_i$
with appropriate hypergeometric function expansions \cite{abramowitz},
yielding near $U = U_i$
\begin{eqnarray} \label{zexpansion}
 A &\ \sim\ &\sqrt{U_i} \ \left( \ {4\over\pi}\ +\
 {z\over 2 \pi} ( 1 - \ln{z\over 32} ) \ +\ {z^2\over 32 \pi}
 ( {3\over 2}\ +\ \ln{z\over 32} ) \ +\ {\cal O}(z^3) \right)
\nonumber \\
 A_D &\ \sim\ & i\sqrt{U_i} \ \left( {z\over 2} \ -\ {z^2\over 32}
 \ +\ {3 z^3\over 2^9}\ + \ {\cal O}(z^4) \right)
\nonumber \\
 \tau_D &\ \sim\ & {i\over\pi} \left( -\ln{z\over 32}\ +\ {z\over 4}
 \ -\ {13\over 256}z^2 \ +\ {\cal O}(z^3)\right)\ ,
\end{eqnarray}

The singular point $U_i$ can be obtained by matching to the one-loop
beta function result, in the perturbative large-$U$ regime, and up to
threshold corrections represented by a constant $c$ of order one, is
\begin{equation} \label{Lambdadefn}
 U_i \approx \Lambda^2 \equiv c\Lambda_{UV}^2 \exp{(i\pi\tau_0)}\ ,
\end{equation}
where $\Lambda_{UV}^2$ is the Wilsonian scale associated with the ``bare''
Lagrangian.  The scalar potential in the dual theory, in the vicinity
of either singular point where the dual theory is weakly coupled, is
thus (to order $|a_D|^2 / \Lambda^2$)
\begin{equation} \label{scalarpotential}
 V\ \approx\ 2 |a_D|^2 \left(|a_M|^2 + |a_{\tilde M}|^2\right)
 \ -\ \pi^2 \ { \left(|a_M|^2 - |a_{\tilde M}|^2\right)^2
  \ +\ 16 |a_M a_{\tilde M}|^2 \over 2\ln{|a_D/16\Lambda|} }\ .
\end{equation}
This is minimized by $a_M = a_{\tilde{M}} = 0$, leaving $a_D$ as
a quantum moduli space.  At $a_D = 0$ the monopoles are massless.

Before introducing $N=2$ supersymmetry breaking mass terms into the
model, it is worth discussing the region of validity of the effective
theory discussed above.  The Wilsonian scale $\mu^2$ must
stay below the scale $U$ where the $SU(2)$ gauge group is broken to
the $U(1)$ subgroup.  Furthermore, the monopole fields must have masses
in excess of $\mu$, since we have cut off the monopoles' contribution
to the $\tau$ function at their mass $\sim (U-U_i)/\sqrt{U_i}$.
(If instead the monopole mass were less than $\mu$, then some of the
gauge coupling's running would be due to loops computed in the low
energy effective theory with internal momenta below $\mu$,
and the $\tau$ function would not depend on $(U-U_i)/\sqrt{U_i}$.)
The vacuum structure of the theories can be obtained from the limit
$\mu \rightarrow 0$, so the effective theory is useful even close to
the point $a_D = 0$ where the monopoles are light.

If we are to extend the $N=2$ model's effective theory to models with
$N=2$ breaking masses these models must continue to exhibit an energy
range with a $U(1)$ gauge symmetry.  As the $N=2$ breaking masses grow
we expect the theories' mass gap to increase (since the cancellations
preventing confinement of the $SU(2)$ gluons will break down).  For
an energy range with a $U(1)$ gauge symmetry to exist, the scale at
which the $U(1)$ gauge boson is confined by the $SU(2)$ gauge dynamics,
and also therefore the $N=2$ breaking masses, must be less than the
$SU(2)$ breaking scale $U$.  The
resulting tractable models occupy a ``ball'' in parameter space around
the $N=2$ model.  At tree level these models possess an $SU(2)$ gauge
symmetry; a light adjoint of fermions; and a light adjoint of scalars,
not necessarily degenerate with the fermions.
We cannot rule out the possibility that the models' qualitative
properties change dramatically as the soft breakings are taken larger
than $\Lambda$ (necessary to recover the QCD limit), although for
small soft breakings we find that the behavior is smooth.

Even though introducing $N=2$ breaking terms lifts the degeneracy of
the moduli space, we can still discuss monodromies of the effective
$\tau$ function in what was the moduli space by coupling a chiral
superfield source $J(x)$ to the adjoint field $\Phi(x)$.
The additional term
\begin{equation} \label{source}
 {\cal L}_J\ =\ \int d^2 \theta\ J \Phi \ +\ {\rm h.c.}
\end{equation}
shifts $\langle a_{\Phi} \rangle$ away from its ($J=0$) vacuum value
when $\langle f_J \rangle \neq 0$.
The additional interaction in (\ref{source}) can of course induce new
$J$-dependent terms in the effective Lagrangian ${\cal L}_{eff}^{J}$.
However, for small $N=2$ breaking masses a background source with
magnitude of order the small breaking mass is sufficient to explore
the monodromies around the original $N=2$ singular points.  In the
case of the gauge kinetic term, the lowest order term induced by $J$,
\begin{equation} \label{Jinduced}
 (\ J^\dagger J \ (D W / \Lambda^3)^2 \ ) \Big\vert_D \ \sim\
 \vert f_J / \Lambda^3 \vert^2 \ F_{\mu\nu}^2 \ ,
\end{equation}
is clearly a subleading contribution near the $N=2$ limit, leaving the
monodromies and singularities of the broken model still controlled by
the $N=2$ original $\tau$ function.  We therefore expect to find light
monopoles or dyons in the effective theory for $\langle a_\Phi \rangle$
near the $N=2$ singular points.

\subsection{Breaking to $N=1$ supersymmetry} \label{sec-2.3}

The $N=2$ supersymmetric model discussed in section~\ref{sec-2.2}
can be broken to an $N=1$ model by introducing a mass term for the
adjoint matter fields \cite{SW,IS}, corresponding in the enlarged model
to allowing $\langle a_m \rangle \neq 0$.
The effective Lagrangian is thus that of the $N=2$ model plus terms,
invariant under the $N=1$ supersymmetry and gauged and global $U(1)$
symmetries of the model, that vanish as the adjoint mass is turned off.
The effective superpotential can therefore receive corrections of the
form
\begin{equation} \label{N=1F}
 \Delta W \ =\ m f(U/\Lambda, \tau_0)\Big|_F
\end{equation}
where $f$ is an unknown function.  This unknown mass renormalization
arises from interactions of $A$ with $SU(2)$ gauge bosons at scales
above $U$.

Introducing this mass term can also give rise to additional
$D$-terms of the form
\begin{equation} \label{N=1D}
\Delta {\cal L}_D \ =\ m^\dagger m\ G \left( {A_D \over \Lambda} ,
 {A_D^\dagger \over \Lambda}, \tau_0,
 \tau_0^\dagger, {MM^\dagger \over \Lambda^2},
 {\tilde{M} M^\dagger \over \Lambda^2},
 {\tilde{M} \tilde{M}^\dagger \over \Lambda^2} \right)\Bigg|_D
\end{equation}
plus higher dimension terms in the gauge fields.
In general $G$ is some complicated unknown function\footnote{
 Functions like $G$ induced by supersymmetry breaking appear in
 $D$-terms of the effective Lagrangian, here and below.  While there
 are some constraints on these functions, for example from weak
 coupling behavior and nonsingularity of the Kahler metric, these
 are usually not sufficient to globally determine the behavior.}.
Note that the corrections in (\ref{N=1D}) to the kinetic energy terms
are suppressed by ${\cal O} (m^2 / \Lambda^2)$,and for small $m$ we
can consider them ${\cal O} (p^4)$.  In this sense these corrections
do not destroy the ${\cal O}(p^2)$ exactness of the $N=2$ solution.
Upon eliminating $f_{A_D}$ from such terms and the canonical $D$-term
for $A_D$, the potential only receives contributions of the form
\begin{equation} \label{VfromWG}
 {1\over 1 + G^\prime\big|_A}\left| {dW\over df_{A_D}}\right|^2
\end{equation}
with $G^\prime$ a function related to $G$.  Now, supersymmetric
vacua satisfy $V = 0$, and $G$ is nonsingular unless the effective
theory completely breaks down.  The extrema of $W$ thus coincide
with the minimum of the potential, and we have
\begin{equation} \label{neq1eqmot}
 \begin{array}{c}
 \sqrt{2}\, a_M a_{\tilde M}\ +\ m \left( d a_U / d a_D \right)\ =\ 0
 \ \ \\ \\ a_D a_M\ =\ a_D a_{\tilde M}\ =\ 0 \ .
 \end{array}
\end{equation}
As Seiberg and Witten found, the potential is minimized for
$\langle a_D \rangle\ =\ 0$ and the monopoles condense with
\begin{equation} \label{neq1minimum}
 a_M\ =\ a_{\tilde M}\ =\ \left( -a_m {a_U}'(0) / \sqrt{2}
 \right)^{1/2}\ .
\end{equation}
The theory possesses a mass gap, since the magnetic $U(1)$ gauge
boson acquires a mass by the Higgs mechanism, and electric charges
are confined.

\subsection{Breaking to $N=0$ with nonzero $f_m$} \label{sec-2.4}

As a first example of soft supersymmetry breaking consider giving a
nonzero vacuum expectation value to the $f$-component of $m$ in the
$N=1$ model.  At tree level this generates an extra $N=1$ breaking
mass term for $a_\Phi$
\begin{equation} \label{fmterm}
 2 {\rm Re}( f_m {\rm Tr}(a_\Phi^2) ) \ ,
\end{equation}
which is renormalized by gauge interactions, through the function $f$
in (\ref{N=1F}), but remains in the effective theory.  In addition
the $D$-terms in (\ref{N=1D}) may generate additional terms
\begin{equation} \label{evalfmterm}
 |f_m|^2 \ G\left({A_D \over \Lambda},
 {A_D^\dagger \over \Lambda}, \tau_0, \tau_0^{\dagger},
 {MM^\dagger \over \Lambda^2},
 {\tilde{M} M^{\dagger} \over \Lambda^2},
 {\tilde{M} \tilde{M}^\dagger \over \Lambda^2} \right)\Bigg|_A
 \ +\ a_m f_m G \Big|_F \ +\ {\rm h.c.}
\end{equation}
Firstly, we note that the unconstrained terms are small
(${\cal O}(p^2)$ or higher) when
$m \ll \Lambda$ and $f_m \ll \Lambda^2$.  Secondly, introducing the
soft breaking mass leaves the lowest dimension gauge kinetic term
unchanged from the $N=2$ model (ignoring not only the higher
dimension terms mentioned above, but also suppressed terms such as
$m^\dagger m(DW)^2/\Lambda^4|_D \sim |f_m^2/\Lambda^4|(F_{\mu\nu})^2$).
We conclude from the singularity structure of $\tau$ that even in the
softly broken model there remain two points at which monopole bound
states become massless.  The soft breaking may generate $a_D$
interactions of unknown sign and similarly contribute to the masses
of the scalar components of the monopole fields.

When $f_m$ is the only $N=2$ breaking parameter, these terms are
suppressed by $f_m/\Lambda$ relative to the bare scalar mass term,
but unfortunately the tree level potential is then unbounded below
as $\langle a_U \rangle \rightarrow -\infty$.  Higher dimension
$D$-terms, contributing for example $|a_U|$ terms to the potential,
determine whether it is truly unbounded or is instead minimized at
some finite $a_U$.  Such unknown terms may however be small compared
to an additional $N=2$ breaking mass generated by $\langle a_m \rangle
\neq 0$.  In the limit $f_m/\Lambda^2 \ll a_m/\Lambda \ll 1$, the bare
masses dominate the corrections to the pure $N=2$ model and we obtain
the potential
\begin{eqnarray} \label{fmscalarpotential}
 V\ \approx\ {-8 \pi^2\over \ln{|a_D/16\Lambda|}  } \left|
     a_M a_{\tilde{M}} - {i \sqrt{2} \Lambda a_m }\right|^2
 \ +\ { 2 |a_D|^2 } \left(|a_M|^2 + |a_{\tilde{M}}|^2\right)
\nonumber\\
 \ -\ {\pi^2\over  2 \ln{|a_D/16\Lambda|}}
 \left(|a_M|^2 - |a_{\tilde{M}}|^2\right)^2
 \ +\ 4 {\rm Im} (f_m \Lambda a_D)
\end{eqnarray}
which is minimized (via (\ref{neq1minimum}), up to terms of order
$f_m^2$ and $a_m^2$) by $a_M = a_{\tilde M} = (i \sqrt{2} \Lambda
a_m )^{1/2}$ and $a_D = i 2 \sqrt{2}f_m^*/a_m$.
We thus obtain an ${\cal O} (p^0)$ solution of a model
with $N=0$ supersymmetry.

We conclude that up to small corrections the vacuum structure of a
model with both a small $N=1$ preserving mass and a small soft
supersymmetry breaking mass is equivalent to that of the pure $N=1$
model.  There thus exist $N=0$ models which are close in parameter
space to Seiberg and Witten's $N=1$ model that exhibit monopole
condensation and confinement.  Unfortunately the terms induced in the
potential by soft breaking from $D$-terms are not determined by the
super-, gauge or global symmetries of the model and thus we draw no
conclusions as to the vacuum structure when
$\langle a_m \rangle \rightarrow 0$.
This theme will recur in the discussion below.

\subsection{Breaking to $N=0$ with a gaugino mass} \label{sec-2.5}

The $N=2$ model can be perturbed directly to an $N=0$, softly broken
supersymmetric model by including a gaugino mass term.  In the
enlarged model such a mass can arise from setting $\beta_\tau \neq 0$
and hence $\langle f_{\tau_0} \rangle \neq 0$.  The effective
Lagrangian is again that of the $N=2$ model plus, potentially, all
additional terms consistent with the symmetries of the model that
vanish as the gaugino mass is switched off.  The lowest dimension
gauge kinetic term is given by the $\tau$ function of $N=2$ and hence
the singularities indicating the presence of massless monopole fields
are unchanged by the soft supersymmetry breaking.  There are however
extra terms generated in the dual theory close to $U_i$, arising from
the terms in the $N=2$ effective Lagrangian when
$\langle f_{\tau_0} \rangle \neq 0$.  In particular, $f_{\tau_0}$
enters through $U_i$'s dependence (\ref{Lambdadefn}) on $\tau_0$;
to second order in $a_z$, with $z \equiv \left(U - U_i \right)/U_i$,
we have
\begin{eqnarray} \label{dualz2potential}
 V &\approx &
  \left|{\Lambda f_{\tau_0} \over 16}\right|^2
 \left( -\ln{|a_z|\over 32}
  + {a_z + a_z^*\over 8}\left(\ln{|a_z|\over 32} + 1\right)
  \right)^{-1}
 \nonumber\\
 && \left|
  8 - {32\sqrt{2}\pi \over \Lambda f_{\tau_0}} a_M^* a_{\tilde M}^*
  - 2 a_z\left(\ln{|a_z|\over 32} - 1/2\right)
  - a_z^*\left( 1 -
     {4\sqrt{2}\pi\over f_{\tau_0} \Lambda} a_M^* a_{\tilde M}^*\right)
 \right|^2
 \nonumber\\
 && - \left|{\Lambda f_{\tau_0} \over 4}\right|^2 \left(a_z
 - {a_z^2 
 \over 16}
    - {4\sqrt{2}\pi \over \Lambda^* f_{\tau_0}^*} a_M a_{\tilde M} a_z
      \right) + {\rm h.c.} \nonumber\\
 && + {\left|\Lambda\right|^2 \over 2}|a_z|^2
      \left(|a_M|^2 + |a_{\tilde M}|^2\right)
 \nonumber\\
 && + {\pi^2 \over 2} \left(-\ln{|a_z|\over 32} + {a_z + a_z^*\over 8}
 \right)^{-1}
       \left(|a_M|^2 - |a_{\tilde M}|^2 \right)^2
\ .
\end{eqnarray}
There may in addition be $D$-terms that vanish in the $N=2$ limit
and give contributions to the superpotential of the form
\begin{equation} \label{ftaudirectD}
 \Lambda^2 H(\tau_0^\dagger, \tau_0, z, z^\dagger)\Big|_D \ .
\end{equation}
Such terms may induce a scalar monopole mass contribution,
for example
\begin{equation} \label{mpolemassexample}
 U_i^\dagger z \Big|_D \sim\ \Lambda^2 f_{\tau_0}^* f_z \ ,
\end{equation}
which on eliminating $f_z$ provide a shift proportional to
$f_{\tau_0}$ within $|f_z|^2$ in (\ref{dualz2potential}).
The resulting cross terms with the monopole fields induce a scalar
monopole mass.  Its sign and magnitude, which would signify whether
they condense, is therefore undetermined by the symmetries.

These terms perturb, in an unknown fashion, the potential
(\ref{dualz2potential}).
One can however read off the gaugino mass in the effective theory,
arising from the $f$-component of the $\tau$ function, as
\begin{equation} \label{gluinomass}
 m_{\tilde{\gamma}_D} \ =\  {m_{\tilde\gamma} \over a_{\tau}}
 \ \sim\ \tau_D^\prime (a_{\tau_0}) f_{\tau_0}
 \ \sim\ -f_{\tau_0} {\Lambda^2 \over a_D}\ .
\end{equation}
The gaugino and dual gaugino masses diverge as $a_D \rightarrow 0$,
which we may interpret as the decoupling of the massive gaugino
fields from the effective theory, as the decreasing monopole mass
drives its region of validity to zero.  If the theory behaves
without singularity as $\langle f_{\tau_0} \rangle \rightarrow 0$,
the contribution of (\ref{ftaudirectD}) to the potential must cause
the vacuum expectation $\langle a_D \rangle$ to become nonzero and
proportional to $\langle f_{\tau_0}\rangle ^l$ with $l<1$, so that
$m_{\tilde{\gamma}_D}$ smoothly approaches zero in the $N=2$ limit.

As before if the gaugino mass is introduced as a perturbation to the
$N=1$ model ($\langle f_{\tau_0} \rangle \ll \langle a_m \rangle$),
then the corrections to the potential from the gaugino mass will be
small and the minima will be that of the $N=1$ model with monopole
condensation.

\subsection{Breaking to $N=0$ with a squark mass} \label{sec-2.6}

Allowing $\langle f_K \rangle \neq 0$ in (\ref{smallLagrangian})
generates a soft supersymmetry breaking squark mass, induced at tree
level by the term $|f_K|^2 |a_{\Phi}|^2$.  Since $K$ has arbitrary
$U(1)$ charge it could not appear in the $F$-terms of the effective
theory, but it does appear in $D$-terms of the form
\begin{eqnarray} \label{effectiveKterm}
 {\cal L}_K &\ =\ &\left[ K^\dagger K\ \Phi^\dagger \Phi\ G_1
 \left(\tau_0, \tau_0^\dagger,
 {\Phi^\dagger \Phi \over \Lambda^2},
 {\Phi^2 \over \Lambda^2},
 {M^\dagger M \over \Lambda^2},
 {{\tilde M}^\dagger {\tilde M} \over \Lambda^2} \right)
 \right. \nonumber \\ && \left.
 \ +\ \ K^\dagger K\ \Phi^2 \ G_2 \left(\tau_0, \tau_0^\dagger,
  {\Phi^\dagger \Phi \over \Lambda^2}, {\Phi^2 \over \Lambda^2},
  {M^\dagger M \over \Lambda^2},
  {{\tilde M}^\dagger {\tilde M} \over \Lambda^2} \right)
 \right]_D \ ,
\end{eqnarray}
with $G_1$ and $G_2$ unknown functions.  The bare squark mass is
preserved up to gauge renormalization, but the induced monopole mass
is undetermined and hence so is the vacuum structure.  The $\tau$
function of the $N=2$ model is again preserved, indicating the
presence of massless monopole states at $a_D = 0$.  The potential
minima when the $N=1$ model is perturbed will still lie close
(for small $\langle f_K \rangle$) to $a_D = 0$ and $a_M$ given by
(\ref{neq1minimum}), not altering the vacuum structure from the pure
$N=1$ case.

\section{$SU(N_c)$ Gauge Symmetry And Matter Fields} \label{sec-matter}

The exact results for $N=2$ supersymmetric QCD with an $SU(2)$ gauge
group have been generalized to the case of an $SU(N_c)$ gauge group
\cite{Ncolor}.  The potential for the adjoint scalars is minimized by
a vev that classically breaks $SU(N_c)$ to $U(1)^{N_c}$.  For large
scalar vev the $U(1)$ couplings can again be calculated in perturbation
theory and shown to be
consistent with the existence of $N$ points on the quantum moduli space
at each of which $N-1$ dyons become massless.  The function $\tau$ is
again exactly specified (up to scaling) by the singularity structure.

Seiberg and Witten have also demonstrated how $N=2$ matter multiplets
can be included in the $SU(2)$ model.  For heavy matter fields, the
$SU(2)$ dynamics are invariant at low energies, except for the
addition of an extra singularity at large $U = m/\sqrt{2}$
corresponding to where a matter field mass vanishes due to its Yukawa
coupling to the scalar vev.  As the mass decreases, the singularity
moves towards the origin of the moduli space and merges with the pure
glue singularities as dictated by the requirement that the dyons fill
out multiplets of the relevant flavor symmetries.  Again the $\tau$
function is exactly determined by the monodromy structure.
These results can also be generalized to $SU(N_c)$ gauge models
\cite{matterN}.

In each of these cases the exact form of the $\tau$ function in the
effective theory is known (up to higher dimensional contributions).
We can break the models to $N=0$ models with squark and gaugino
masses by promoting the bare couplings $\tau_0$ and $K$ to the
status of chiral superfields and allowing them to acquire nonzero
$f$-component vevs.  As for the $SU(2)$ theory, the $D$-terms will
generate unknown contributions to the potential, but the gauge
multiplet's kinetic term is given exactly by $\tau$.
If the soft breaking masses are small relative to an $N=1$ preserving
adjoint matter mass then the models are pinned near the singularities
and the monopoles condense.

Among these models is the particularly interesting case of an
$SU(3)$ gauge symmetry and $N=1$ matter multiplets in the fundamental
representation.  We may choose to introduce an $N=1$ supersymmetry
preserving mass for the adjoint matter field and soft supersymmetry
breaking masses for the gauginos and fundamental representation
scalars.  At tree level these masses can be raised until the fields
are integrated from the theory, leaving QCD with quarks.  Although we
cannot explicitly take this limit in the effective theory, since the
light degrees of freedom there are not those appropriate to QCD, we may
begin the interpolation from the $N=2$ result towards QCD.  The behavior
of the theory is that monopoles condense and the quarks are confined.
This picture's consistency with the t'Hooft--Mandlestam picture
\cite{dualqcd} of quark confinement in QCD suggests that the models
are smoothly connected.

\section{Discussion} \label{sec-discussion}

We have embedded the $N=2$ and $N=1$ supersymmetric models possessing
exactly calculable results within a larger $N=1$ model, which in
certain limits induces soft supersymmetry-breaking masses in the
original theories.  This procedure generates new contributions to
the effective Lagrangian, which unfortunately holomorphy and $U(1)$
symmetries do not completely determine.  Thus the information
we are able to extract about the behavior of corresponding $N=0$
models is limited.  However, at leading order in the breaking masses
the gauge kinetic terms are still completely determined, leaving
unchanged the $\tau$ function of the $N=2$ and $N=1$ models.
Since the singular behavior of $\tau$ is consistent with the original
ansatz that a monopole becomes massless at each of the singular
points, we expect that these are still the only singularities in the
low-energy effective Lagrangian of the nonsupersymmetric models.
If the soft supersymmetry breaking masses are induced as sufficiently
small perturbations to the $N=1$ preserving model, then monopole
condensation and confinement are preserved and seen not to depend on
the presence of exactly massless gauginos or squarks, nor on exact
supersymmetry degeneracies.
While we have encountered some limitations in applying Seiberg and
Witten's techniques to nonsupersymmetric models, the analysis
nevertheless reveals some details of condensation and confinement.
Our results are consistent with the supersymmetric models being
smoothly connected to nonsupersymmetric QCD-like models.

\begin{flushleft} {\Large\bf Acknowledgments} \end{flushleft}

The authors are grateful to Luis Alvarez-Gaum\'e, Jacques Distler,
Mans Henningson, Greg Moore, Nati Seiberg and Matt Strassler for
useful discussions and comments.  This work was supported in part
under United States Department of Energy contract DE-AC02-ERU3075.
SDH acknowledges the hospitality of the Aspen and the Benasque
Centers for Physics, where some of this work was performed.

\newpage

\end{document}